# Executable Modeling with UML
## - A Vision or a Nightmare? -

Bernhard Rumpe

*Software & Systems Engineering*

*Munich University of Technology*

*D-80230 Munich, Germany*

*Tel: +49-89-45364800, fax-4738,*

*Rumpe@in.tum.de*

### Abstract

*Extreme Programming is the most prominent new, light-weight (or agile) methods, defined to contrast the current heavy-weight and partially overloaded object-oriented methods. It focuses on the core issues of software technology. One of its principles is not to rely on diagrams to document a system. In this paper, we examine what properties a modeling language like UML must have in order to support the Extreme Programming approach effectively. In particular, we discuss how such a diagrammatic programming language must look like to replace a textual programming language and what benefits and problems such an approach may bring..*

Keywords: UML, extreme modeling, visual programming language, executability.

## 1. INTRODUCTION

Extreme Programming (XP) [1] is a light-weight methodology for small and medium-sized teams developing software with rapidly changing or enhancing requirements. XP is an explicit reaction to the complexity of today's modeling methods like the Unified Process [12], the OPEN Toolbox of Techniques [7], or Catalysis [3]. XP primarily tries to focus on the best practices of software development, and contrasts strongly the heavy-weight and partially overloaded object-oriented methods by its simplicity.

In all engineering disciplines nowadays, software engineering excluded, there exists an established engineering process to develop a system, which is accompanied by a number of suited modeling description techniques. Software engineering, being a rather new field, has not as yet established any clear methodical guidance or a fully standardized modeling notation. The XP approach does not try to create a detailed software engineering process, but to focus mainly on programming since it is "fun for the programmer" [1]. The current success of this approach suggests that software engineering may either be completely different from other engineering disciplines or that the software engineering discipline simply isn't mature yet. The former has its justification in the fact that software is totally immaterial, whereas all other engineering products have a physical manifestation. Therefore, it is in principle far easier to change already existing software, even if it had been shipped and installed millions of times.

One of the distinct features of XP is the lack of any documentation whatsoever, except for the code itself. This is a contraposition to the modeling techniques like the Unified Modeling Language (UML) [2], [15] which strongly focus on documentation. XP takes an extreme position there, not even documenting the architecture of the system. Often, it is very difficult to extract the overall structure, behavior or interactions with the environment from the code. The code is a rather detailed and fragile representation of the system's tasks. Even though the code contains all necessary information about the system, this information is often burdened with details and it is tedious to extract the aspects one is interested in. Therefore, it would be useful to have a more compact system representation. The UML does provide a number of notations that are suited for this purpose. However, the tools so far are not capable of supporting UML in such a manner that it can be well-integrated with the approach of Extreme Programming.

This paper explores which kind of concepts, tools and techniques are needed to make UML suitable for an "extreme modeling" approach. The XP approach is basically a programming approach, but replacing the underlying programming language by an executable version of UML.

In Section 2, we explore techniques and tools needed for the UML to support the extreme modeling approach. In Section 3, we examine in detail UML's actual version and how far it supports extreme modeling. In Section 4, we are going to examine description techniques of the UML that are executable or of use for testing. In Section 5, we discuss our vision of the UML to support the extreme modeling process in the future. In Section 6 we finally discuss drawbacks and changes of such an approach.

## 2. PROPERTIES NEEDED FOR UML TO SUPPORT EXTREME MODELING

UML is, as its name states, a modeling language. The OMG, standardizing UML, explicitly wants it to remain independent from methodical issues. Therefore, the language UML is usable for a variety of purposes. Unfortunately, current tool support for UML is definitely insufficient for many of the purposes for which UML can be used. In this section, we are going to explore what UML needs in order to be able to replace a programming language in the XP approach. We have identified the following six important issues:

1. UML needs to be fully expressive,
2. UML needs to be a more compact notation than an ordinary programming language,
3. UML needs an effective translation into efficient code,
4. UML needs support for testing,
5. UML needs a simple and usable module concept,
6. The tool support must be adequate.

Let's explain the above issues: if UML is going to replace an ordinary programming language, it needs full expressiveness. This has two different flavors; from theory, we know that full expressiveness means Turing computability. All ordinary programming languages, be they object-oriented, structured or functional, have basically the same power of computability, namely, the power of the Turing machine. Therefore, the executable version of UML must also provide the possibility of defining any computable function.

For practical purposes, it is also important that the modeling language we use is expressive enough to describe the connection to the graphical user interface and to the operating system, as well as e.g. distribution aspects, just as today's ordinary programming languages provide by means of appropriate libraries.

One of the advantages of a diagrammatic language will surely be that it can describe structural and behavioral issues in a way more compact and easy to survey than ordinary textual programming languages can. In the case that the UML becomes fully expressive, this advantage has to be maintained.

An effective translation into efficient code is the most critical issue for the Extreme Programming approach. The effective translation means that a compiler from UML to executable code is fast and reliable. The generated code itself must be fast and robust in order to be accepted by users and programmers. Of course, each translation of higher order concepts into a less rich language generates some overhead, but this overhead must be efficient, small, and not too time-consuming. This situation can be compared to object-oriented languages versus structured languages, where the concept of objects generates some overhead which is today efficient enough to have no serious impact on computation time.

In Extreme Programming, the system code and the testing code are both written in the same programming language. In UML, it may be of interest to identify a subset of the language dedicated to specification of tests and to check them during the run-time against the executable model. Moreover, like in any other language, it is important to have a module concept that allows us to encapsulate internal implementations. Only a clear concept of interfaces between the modules and concepts used in the language allow us to distribute and thus parallelize labor among the team.

Finally, tool-support for the techniques discussed above are crucial to the whole approach. Extreme Programming and extreme modeling as such heavily rely on appropriate tool support.

## 3. UML TODAY

Based on the needs identified in the last section, we are now going to identify which properties are already satisfied by UML or how to reshape UML to make it usable for the XP approach.

**3.1. Is UML fully expressive?**

UML consists of several parts. Looking at the diagrammatic part only, UML is not expressive enough to describe each computable function. However, this is not fully clear today because UML semantics is not as precisely defined as necessary to clarify this question. If we also regard the Object Constraint Language (OCL) [21] we find a rich textual language to describe properties of our system. Even though the OCL is a first-order language, it is very much in the spirit of being executable. UML coupled with an executable sub-language of OCL will be expressive enough to describe each possible computation function.

The question of how to access graphical user interfaces and operating systems, and other issues of this kind will probably be solved in the same way as in ordinary programming languages. This means, we need to provide modeling libraries especially suited for these issues, that have a hard-coded implementation. Providing modeling libraries for reusable code would, of course, always be useful.

### 3.2. Is UML more abstract than an ordinary programming language?

If used in a version similar to the current UML 1.3, we can clearly say: Yes, it is more compact than any existing ordinary programming language. This comes from the fact that UML offers a number of higher-level modeling concepts, allowing a rather compact description of certain properties of the system. Describing such a property in ordinary programming language needs considerably more lines of code.

Besides being more compact, abstraction has a second, closely related advantage. A notation is more abstract, if it allows to disregard certain details of the implementation. In UML e.g. it is possible to draw an association without specifying how this association is implemented. Instead, a tool may decide what is the best way to implement the association if this is at all necessary. This may depend on the usage of associations, as well as on the used heuristics to produce code in the tool. However, the implementers should not care about this.

### 3.3. Is there a UML compiler into efficient code?

Not today. For a large part of UML diagrams today, exists not even a single translation into code at all, though many of the UML diagrams look like being executable (see also Section 4.).

Today's tools mainly focus on translating class diagrams into codes and vice versa. This technique is called round-trip engineering and reminds on early stages of compilers from Pascal, Basic or Cobol into assembler code. In these early phases, people had less confidence in their compiler, so they generated assembler code which could be viewed and changed by hand, if necessary, before generating object code. If history repeats, round-trip engineering will vanish sometime and UML models will directly be translated into object code without any intermediate programming language notation.

### 3.4. Is UML suited for testing?

Yes, UML clearly has potential to model tests as well as the executable code. Sequence diagrams and collaboration diagrams both are exemplaric notations that allow especially to describe expected behavior and expected changes of the object structures. Thus both notations are well-suited to describe tests for the system. We already know that UML provides a more compact code description than an ordinary programming language does. But, the tests can also be described in a more compact way, giving rise for specification-based testing. This is, of course, only feasible if appropriate tools for this kind of testing actually exist.

### 3.5. Does UML have an appropriate module concept?

The concept of a module is the basic constituent for programming in the large. Only appropriate module concepts allow to structure the work in a team leading to a parallelization of work, therefore, to a smaller time-to-market. As modules interact, a clear concept of interfaces between modules of the software system becomes indispensable.

In object-oriented programming languages, the concepts of class and package fulfill the requirements for modules. Classes provide a name space that allows hiding of the implementation details and having a clear interface to clients of the class. However, classes are rather low-level. If a system has hundreds of classes, it is necessary to have appropriate structuring mechanism beyond single classes. Therefore, UML uses its package concept to structure classes. This package concept is rather powerful but, so far, not fully elaborated and understood. It seems to be inappropriate to some extent due to the lack of clear interfaces between the packages, even if the package concept can be probably adapted to describe interfaces between modeling elements. Packages do have a name space. However in their current usage, it is not possible to define package interfaces explicitly. Such an interface could be used to explicitly export parts of a package to another one, i.e. to a part of defined classes or type definitions. Due to the lack of sufficient, precise specification or even of a powerful tool support for UML packages, it is not clear so far whether packages will fully support the possibility of separation of concerns, and therefore, localizing changes of the software to smaller parts.

**3.6. Is there adequate tool support for UML as extreme modeling notation?**

Today UML does have an extensive tool support. These tools, however, focus strongly on editing and version control support. Apart from generating code frames and supporting round-trip engineering with class diagrams to a limited extent, there is no extensive tool support today for the UML. Besides the main-stream UML tools, there are a number of tools originating from earlier efforts that have at least some of the concepts mentioned above. Some examples are ROOM [20] or as called today UML-RT, Statemate based on the StateChart formalism [5], and a number of more academic tools like AUTOFOCUS [8].

To summarize - UML has potential for a high-level programming language. However, a number of minor flaws need to be fixed, and especially an executable subset of UML needs to be identified. The main obstacle remains the lack of appropriate tool support for an executable UML that goes far beyond class diagrams and that is reliable enough.

## 4. CAN UML BE EXECUTED?

UML has many interesting facets, some of them already investigated in the last section. Here, we discuss which of the UML models can be executed.

Apart from the question whether UML **should** be executable, that will be discussed in Section 6, let us discuss the question of whether UML **can** be executable. The answer is: yes. There is a large subset of UML modeling techniques that can actually be animated. Among them are the UML class diagrams, StateChart diagrams, the Object Constraint Language, activity diagrams and sequence diagrams.

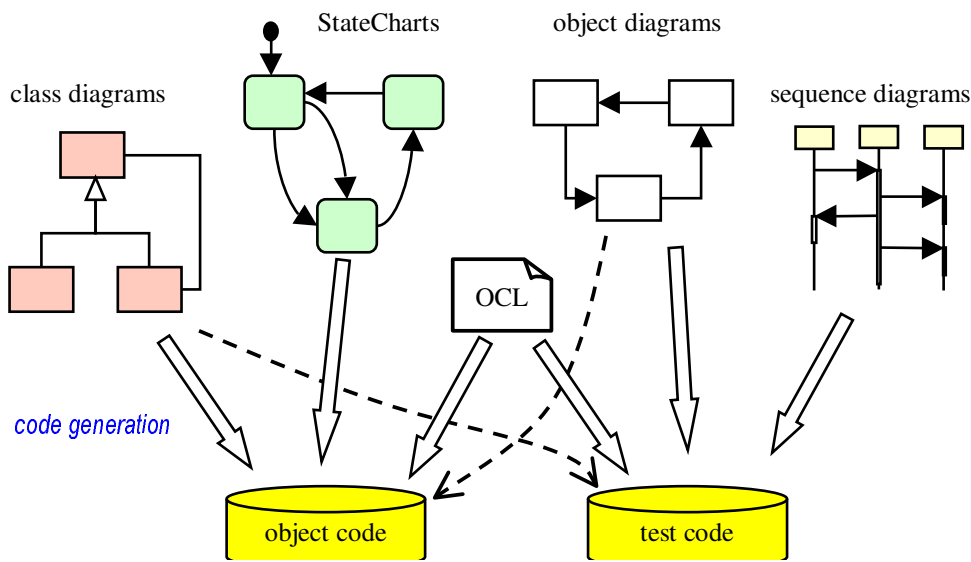

### 4.1. Class Diagrams

How to generate code out of class diagrams is already well-understood. Also, the reverse direction, how to generate class diagrams out of the code is implemented in many UML-based tools today to allow round-trip engineering. From the mathematical point of view, these two mappings are not isomorphisms, which means that mapping class-diagrams to code looses information. It especially looses information about weak or strong aggregation and some information about its associations. Mapping code to class diagrams also looses information, since e.g. method bodies are not (officially) represented in the class diagram. This is due to the fact that a class diagram is an abstraction of the real system that mainly deals with structure.

Besides class diagrams, there are also object diagrams that deal with structure. Whereas class diagrams define and constrain potential structures of a system, an object diagram defines an actual structure of objects in a system in a certain situation. Therefore, object diagrams operate on the instance level. This makes it easy to generate code out of an object diagram, e.g. a code that builds up the object structure. However, there is only poor tool support today for this kind of diagrams.

### 4.2. StateChart Diagrams

In the embedded systems area where control states are a major issue, one is acquainted with the description of complex controlling systems using state machines. In UML, the corresponding notation called StateCharts can also be used to describe behavior of single objects. According to David Harel, StateCharts are the engine of the UML model [6]. StateCharts are a more descriptive and elegant way to describe automata or state machines, and StateCharts are therefore strategies of how to execute the behavior. Thus, it is not surprising that StateCharts are executable and a number of tools, among them also Statemate, demonstrate their efficient executability. Even though and syntax of StateCharts, as used within UML, are syntactically and semantically, as well as from methodical usage slightly different from the original StateCharts definition [5], StateCharts in UML still remain executable, as e.g. the tool ROSE-RT demonstrates.

### 4.3. OCL

The UML diagrams are a powerful set of techniques to describe different views on a software system. Generally though, they are not capable of describing every possible property. The textual constraint language OCL, therefore, has been added to the UML in order to describe properties not to be conveniently captured by diagrams. OCL claims to be a first-order specification language and it has actually the concept of quantification included. However, a close examination of OCL shows that almost all concepts are executable and OCL can be regarded to a large extent as a functional programming language like ML [17], Haskell [9] or Gofer [13]. An interesting aspect concerns the quantifiers. There is an existential and an universal quantification. Assuming there exists a class `Person` with an attribute age, we can write the following OCL constraint:

```
Person.allInstances->forall(p | p.age > 4)
```

For a specification language, OCL does have an unusual syntax. The above invariant ensures for all existing persons in a system the attribute `age` greater than `4`. As in any point of time, each running system has a finite set of existing `Person` objects, this quantification is always finite. Therefore, OCL does not possess the power of first-order logic, but of a propositional logic only and the OCL constraints can be checked at runtime.

Any usage of universal and existential quantification ranges over finite sets only. - With one exception: in UML 1.3 both quantifiers may be used over basic datatypes like Integer or String. The UML 1.3 specification does not define exactly what this really means. We basically have two possible interpretations at hand. In one interpretation, the following constraint

```
Integer.allInstances->forall(x | x <> 5)
```

is not necessarily false. In one interpretation the expression `Integer.allinstances` corresponds to all existing integers in a system snapshot. This set is finite and the constraint above simply states that integer `5` is not assigned to any variable of the system at all. In the other interpretation, `Integer.allinstances` is the actual set of all integers, and the above statement wrong. This means we have infinite quantifications at hand to describe properties, but of course, such quantification normally cannot be executed.

Even if we assume that OCL is completely executable, we still have the problem that describing a constraint does not tell us how to establish it. For example, if we describe a post-condition or an invariant with OCL, there is no automatic way to generate code from it that is capable of actually establishing the post-condition or the invariant.

Because of this big difference between checking a constraint and establishing it, UML 2.0 will probably be extended by an action language. So long, it is not quite clear what the action language will look like, but its purpose will be to describe behavior. We expect the action language to be basically executable and to allow us to describe the actions an object performs when receiving a stimulus.

### 4.4. Activity Diagrams

Apart from the discussed kinds of diagrams, UML contains a number of additional diagrams, which play a less important role in the UML of today. One of them is the newly introduced activity diagram. Even though activity diagrams are presented as a specialization of StateCharts, they actually extend the properties of StateCharts allowing us to describe several concurrent threads of execution. Depending on their actual semantics, activity diagrams are a mixture between dataflow diagrams, which have already been present in OMT [19], and Petri-Nets [18]. For both, dataflow and Petri-Nets, exist tools to execute them, so we can expect with high certainty that activity diagrams will also be executable.

### 4.5. Sequence Diagrams, Collaborations

The diagrams discussed so far, are capable of describing complete sets of possible structures or possible behaviors. In contrast, sequence diagrams and collaboration diagrams describe the system on an instance level. Both diagrams show one possible, exemplaric behavior. This is perfect to model stories and to discuss them with users, but having a finite set of sequence diagrams available, we cannot generate complete code out of it. Regarding code generation, this flaw has been addressed by extending sequence diagrams with techniques for describing alternatives, iteration and repetition. Some of these techniques are already available in UML sequence diagrams. Deeper considerations have been published in [14] and partly included in the Message Sequence Diagrams (MSC) as standardized by ITU [11]. If we extend further sequence diagrams, i.e. by concepts like actions in the activity bars, then we will probably have sequence diagrams that allow to generate complete code. However, these techniques make sequence diagrams more complicated, and we have plenty other description techniques at hand which allow code-generation.

The exemplaric nature of sequence diagrams on the other hand, offers an opportunity to specify test cases. A sequence diagram can be interpreted as a test-driver, sending a sequence of stimuli to an initially created object structure. The sequence diagram is furthermore capable to describe the responses of the object structure under test as well as internal flow of messages. Thus it can be expected that sequence diagrams will be used for coding test-drivers and expected message flows.

## 5. UML APPLIED IN THE EXTREME MODELING APPROACH

XP pretends to use a number of common sense principles and practices to rather extreme levels. In this section, we are going to examine these practices in detail and see how our vision of UML supports them. We refer to the best practices given in [1]:

- XP wants an early, concrete and continuing feedback through short development cycles. This is even more true when we have a more compact modeling language at hand, which allows us to describe more effectively the properties of the system. We achieve even shorter development cycles than an ordinary programming language could enable.

- XP relies on an incremental programming approach which allows to come up quickly with an overall plan that is continuously evolved during a project's life. Having a notation such as class diagrams at hand, the incremental planning is even easier than only working with a programming language. Furthermore, evolution of a system is also strongly supported as refactoring techniques [16], [4] strongly rely on class diagrams. So both incremental planning and evolution are even better supported by UML than by some ordinary programming languages.

- XP relies on its ability to flexibly schedule the implementation of new functionality directly responding to enhancing or changing business needs. This is a methodical issue, there is no reason why the use of UML as programming language should not better support this than the current programming languages.

- XP strongly relies on automated tests written by programmers and customers to ensure and monitor the progress of development in order to catch the facts as early as possible. Using an ordinary programming language, the program code and the testing code both have to be written in the same language. This is feasible, of course, but having specialized notations for description of the system and of its tests makes it even easier. As discussed in the previous section, we found a number of description techniques, i.e. class diagrams, StateCharts and a subset of OCL, to be well-suited for high level executable modeling. Because of their exemplaric nature, sequence, collaboration, and object diagrams are especially suited to describe tests on an instance level. For example, we can use object diagrams to describe a start situation for a method call and the structural part of its post-condition, namely, the final situation for that method call. The interactions happening during this method can be described by a sequence diagram or a collaboration. These kinds of diagrams are specifically suited to describe tests and test situations, tools could graphically show where a test situation has been violated. This is easy to grasp when the test is broken. XP expects tests not only to be written by programmers but also by customers. Some customers can describe tests in diagrams more easily than describe tests as code. Needless to say, we need a good set of tools supporting diagrams that we are going to use in the XP approach, not only for the purpose of testing.

- XP relies on communication, tests and source code without further documentation to communicate system structure and intent. Of course, it is easier to discuss on precise and abstract pictures than relying only on textual source code. If the source code is partly replaced by UML diagrams, then documentation and code again coincide without having much redundancy - a goal that XP tries to achieve. The more sophisticated and higher level concepts a programming language has, the more compact the notation is, the easier it is to grasp the system structure and intent. We expect from UML as a programming language to be of high advantage regarding the understanding of and the communication of a system.

- XP aims at evolutionary design as long as the system is in use. This is a highly critical point because systems that are unstable tend to change their functionality as well. Unfortunately tests cannot capture the possible behavior totally. Thus, it may happen that certain subtle changes of system functionality are not detected by tests. Of course, this is still possible with the extreme modeling approach, but we hope that using a more compact, higher-level language better assists maintenance and evolution of an existing system than a lower-level programming language does. In particular, the program parts to be adapted in a task might be less distributed in the code and therefore easier to grasp and overview.

- In XP the skills of programmers play an important role. XP tries to match short-term instincts of programmers with long-term interests of the project. Using an extreme modeling approach means that the programmers need some skills and specially some interest in high-level modeling with UML instead of using an ordinary object-oriented or structural programming language. As we still do not have sufficiently powerful tools to support UML as a high-programming language, we cannot have programmers with that skill. We expect that as soon as such tools exist, a great deal of programmers is interested enough to learn the skills necessary to use them. But this is a kind of paradigm shift which is equally difficult or even more difficult than the shift from structural programming like using programming language C to an object-oriented programming language like C++.

In summary, had we a sufficient, useful tool support for a subset of UML as a high-level programming language, the extreme modeling approach as discussed above would be a natural evolution of Extreme Programming process as introduced in [1]. It would have advantages over XP in many common sense principles and practices that XP already relies on today.

## 6. PROBLEMS WITH EXECUTABLE UML

Executing a specification or a specified model has both benefits and drawbacks. As discussed some benefits are the early feedback for developer, experiencing his model's actual behavior. Another important advantage is the fact the earlier you get the code running, the more market you have. Certainly, generating code out of a model gives distinctive time and market advantage. On the other hand such an approach suffers from serious drawbacks.

## 6.1. Efficient specifications

Known from the early efforts when defining executable specification languages, the modeler tends to focus not only on the properties he is modeling, but also on efficiency of the execution. It is hard enough to specify properties concisely and accurately, and it becomes even harder if we are simultaneously concerned with efficiency. Such experience has been made with algebraic specification languages however, and it is not clear whether these considerations also hold for modern diagrammatic languages like UML.

Executable specification of functionality such as sorting are harder to read and understand than abstract and compact property specifications. As OCL is conceptually similar to algebraic specification languages, at least OCL-specifications will suffer from efficiency considerations. On the other hand this problem will be less dramatic for other UML notations, as those mainly deal with structure (class diagrams), or are efficiently executable by their nature (StateCharts).

## 6.2. The Problem of Over-Specification

Our examination has shown that a large sub-language of UML can be executable. The question remains: shall UML be an executable, high-level programming language?

Today, UML based tools often force developers to specify details unknown at the moment, or details they wanted to be left open. This is well-known as the problem of over-specification and will surely become worse if the tools are going to head towards being high-level programming language compilers. Looking at SDL, which had a similar fate starting of as specification language in the telecommunication area, ending up as a high-level programming language, we will probably witness UML making a shift from a property-description language to an executable language, used mainly for programming.

Repeating earlier arguments, it is doubtlessly useful to have an immediate simulation of your model at hand; nevertheless during modeling, especially architectural or requirements modeling, the possibility to under-specify unwanted or unavailable properties is highly important. Therefore, it is useful in general to have both an executable sub-language of UML and a highly non-executable sub-language that allows us to specify system properties declaratively. High-level UML specifications must then be transformed into low-level detailed executable models by adding details and refactoring the models. For example, whereas we need for execution purposes mostly one StateChart for each class during the specification, it might be of great use to have several abstract StateCharts that describe parts of behavior of a class from different points of view, and that are merged into one, more detailed StateChart during the development process. As we know, this kind of merging cannot be done automatically, it needs methodical assistance beyond today's tools that mainly allow drawing diagrams and generate code out of them.

## 6.3. Focus on the target language

When using code generators that map UML to a target language, the semantics of the target language as well as its notational capabilities tend to become visible on the UML level. For example, missing multiple inheritance in Java may restrict executable UML to single inheritance as well. Furthermore, the language internal concurrency concept, message passing or exception handling may impose a certain dialect of executable UML. This proliferates UML dialects as semantically incompatible. In particular it will not (easily) be possible to transfer UML models from one target language to another.

### 6.4. UML in the early phases

Executable UML will be useful for a number of projects. However in many other software development approaches, there are early phases with explicit requirements analysis, specification of the system functionality and development of an explicit architecture of the system. Executable UML will have a number of deficiencies to describe this artifacts. We already mentioned the problem of over-specification and the dependency on the target language. Furthermore, an executable UML will not cover the full UML 1.3 as it is today. For example, we do not expect use cases to become an executable notation, even though there are attempts to generate code from them.

As a consequence, executable UML should be a subset of a larger UML that is capable to assist the developers in the early phases. This extended version of UML relaxes certain restrictions of executable UML, uses more diagrams, and offers a number of high-level concepts within the mentioned diagrams, that cannot be executed.

## 7. CONCLUDING REMARKS

This paper discusses how and whether to integrate modeling techniques that a language like the UML offers in an executable form with the Extreme Programming approach. We have called the result "executable modeling" based on the idea to replace ordinary programming by high-level modeling.

The basic idea of executable modeling is to replace the programming language on which XP heavily relies on by a high-level executable modeling language. We examined the UML potential for this approach and found that it would support most of our needs. However, today's tool support is poor and insufficient. As long as tools do not fully support our needs, the vision in this paper will only remain a vision. However, looking at present situation of UML tools, it seems rather likely that the fate of the Unified Modeling Language will be similar to the fate of SDL [10], which also started as high-level description language in telecommunication area and ended up as a high-level programming language.

Although the development of an executable UML-version has a number of advantages, there are also a number of drawbacks. Most critically, it has to be made clear to software developers, that in the early phases a non-executable UML should be used. There should be no emphasis on efficiency or completeness of the models. Instead the models should be abstract and focused on the information they are intended to describe.

To conclude, there is a necessity for a non-executable as well as an executable version in the UML language family that both will co-exist and be used where appropriate.

## Acknowledgements

The author wishes to thank Kent Beck and Jutta Eckstein for fruitful discussions on XP, also Wolfgang Schwerin for his comments. This work was partially supported by the Bayerische Forschungsstiftung under the FORSOFT research consortium and the Bayerisches Staatsministerium für Wissenschaft, Forschung und Kunst under the Habilitation-Förderpreis program.